\documentclass[]{article}
\usepackage{color}
\usepackage{amsmath,amsfonts,amssymb}
\usepackage{graphicx}
\usepackage{authblk}
\usepackage{url}
\usepackage{textcomp}
\usepackage{caption}
\usepackage{appendix}
\usepackage{hyperref}
\usepackage{gensymb}
\hypersetup{hidelinks}
\usepackage{geometry}
\usepackage{xcolor}

\title{Long-term absolute frequency stabilization of a hybrid-integrated InP-Si$_3$N$_4$ diode laser}

\author[1,$\dag$]{Albert van Rees}
\author[1,2,*,$\dag$]{Lisa~V.~Winkler}
\author[3]{Pierre Brochard}
\author[4]{Dimitri Geskus}
\author[1]{Peter~J.~M.~van~der~Slot}
\author[2]{Christian~Nölleke}
\author[1]{Klaus-J.~Boller}

\affil[1]{Laser Physics and Nonlinear Optics, Faculty of Science and Technology, MESA+ Institute, University of Twente, Enschede, the Netherlands}
\affil[2]{TOPTICA Photonics AG, Gräfelfing, Germany}
\affil[3]{Silentsys SAS, Le Mans, France}
\affil[4]{Chilas BV, Eindhoven, The Netherlands}
\affil[$\dag$]{Both authors contributed equally to this work.}
\affil[*]{Corresponding author: l.v.winkler@utwente.nl}

\begin{document}

\maketitle

\begin{abstract}
Hybrid integrated diode lasers based on combining semiconductor optical amplifiers with low-loss Si$_3$N$_4$-based feedback circuits enable great laser performance for advanced photonic circuits.
In particular, using high-Q Si$_3$N$_4$ ring resonators for frequency-selective feedback provides wide spectral coverage, mode-hop free tuning, and high frequency stability on short timescales, showing as ultra-narrow intrinsic linewidths.
However, many applications also require long-term stability, which can be provided by locking the laser frequency to a suitable reference.
We present the stabilization of a hybrid-integrated laser, which is widely tunable around the central wavelength of 1550~nm, to a fiber-based optical frequency discriminator (OFD) and to an acetylene absorption line.
By locking the laser to the OFD, the laser's fractional frequency stability is improved down to $1.5\cdot10^{-12}$ over an averaging time of 0.5~ms.
For absolute stability over longer times of several days, we successfully lock the laser frequency to an acetylene absorption line.
This limits the frequency deviations of the laser to a range of less than 12~MHz over 5~days.
\end{abstract}

\section{Introduction}
Chip-integrated and frequency-stable diode lasers tunable to various wavelengths in the infrared or visible range are central for a wide range of applications, such as scalable quantum technology~\cite{Niffenegger_2020nat,Mehta_2020nat,Mahmudlu_2023np}, integrated optical atomic clocks~\cite{newman_2019o}, signal processing~\cite{marpaung_2019np, Botter_2022sciadv} and data transmission~\cite{lin_2018pj, Zou_2020ol, Dass_2022jlt}.
A high short-term stability, also expressed as a low intrinsic linewidth, can routinely be achieved with diode lasers in a chip-integrated format, by extending the cavity length with a low-loss waveguide feedback circuit~\cite{Tran_2019apl, Tran_2020jstqe, boller_2019Phot, Fan_2020oe}.
An alternative is using laser-external high-Q resonators for line-narrowing at the predefined frequencies of standard Fabry-Perot, DFB and DBR diode lasers through self-injection locking~\cite{Jin_2021np, Li_2021ol, Siddharth_2022apl, Corato_2023np}.
To also achieve high long-term frequency stability, active stabilization to highly stable frequency references is required.

When considering bulk lasers, gas lasers such as helium-neon lasers are traditionally well-suited for long-term stability, allowing frequency stabilization with drifts of only around $2\cdot10^{-8}$ over several months~\cite{ciddor1983two, rowley1990performance}.
More recently, enormous progress has been achieved with frequency stabilization of diode lasers, enabling similar performance~\cite{arnold1998simple, martin2016external, Krause_2020ao}.
Furthermore, the lowest fractional frequency instabilities have been achieved using state-of-the-art bulk ECDLs~\cite{Herbers_2022ol} and fiber lasers~\cite{Matei_2017prl}.
Notably, a stability of $4\cdot10^{-17}$ up to 10~s was reached by stabilizing a fiber laser to an ultra-stable Fabry-Perot cavity~\cite{Matei_2017prl}.
On the other hand, a waveguide-based diode laser has recently exceeded the short-term frequency stability of high-performance fiber lasers, due to feedback from an ultra-high-Q integrated resonator~\cite{Li_2021ol}. 
Such chip-sized lasers enable seamless integration into mass-produced photonic circuits.
Long-term active stabilization of these integrated diode lasers would be key for repeatability and reliability for a wide range of photonic applications.

To enable long-term stability over multiple hours or even longer, integrated diode lasers require additional properties.
First, the laser should not make any mode hops in the required time frame, as they are challenging to correct for by electronic stabilization.
While operation with individual chips aligned on stages is sufficient for investigating the short-term stability~\cite{Jin_2021np, Li_2021ol}, temperature drifts or acoustic perturbations are prone to cause mode-hops in such a configuration.
To avoid this, the passive stability of the entire waveguide circuit needs to be improved through integration of the feedback chip with the diode laser.
Second, to provide electronic feedback for stabilization, the laser must have a wavelength tuning mechanism with sufficient bandwidth and range.
For also accessing any target wavelength within the laser’s gain bandwidth, specifically single lines of absolute reference absorbers~\cite{Cherfan_2020oe, Zhang_2020lpr, Krause_2020ao}, a suitable diode laser concept has to be chosen.
Diode lasers that use feedback from Bragg waveguides for spectral selection~\cite{numata_2010OE, Liu_2021optica, Guo_2022sa} offer only very limited tunability.
Lasers where frequency noise reduction is based on self-injection locking from microring resonators~\cite{Jin_2021np, Li_2021ol, Siddharth_2022apl, Corato_2023np} offer wider tuning, however, this linewidth narrowing method requires careful adjustment of the feedback phase for constructive interference inside the laser cavity~\cite{Corato_2023np}.
On the other hand, diode lasers employing Vernier-filter based extended cavities provide wide tuning over the entire gain bandwidth~\cite{Guo_2022apl} and enable wide mode-hop free tuning over absorption lines~\cite{VanRees_2020oe}.

Active stabilization of such a widely-tunable laser based on semiconductor integrated photonics has recently been demonstrated with a best frequency stability of $2.5\cdot10^{-13}$ at 1~s integration time\cite{Stern_2020ol}.
An even better frequency instability of $10^{-14}$ below 1~s was achieved with a heterogeneously integrated DBR-laser, self-injection locked to a spiral resonator, and actively stabilized to a lithographically fabricated Fabry-Perot cavity~\cite{Guo_2022sa}.
However, such stabilization schemes have only been demonstrated over intermediate time scales up to several minutes.

Here we demonstrate the absolute frequency stabilization of a hybrid-integrated, widely tunable diode laser over several days.
The laser is formed by hybrid integration of an InP diode amplifier and a Si$_3$N$_4$ feedback circuit.
For the free-running laser, we observe mode-hop free operation over several days, achieved by electronic, thermal and optical packaging of the laser in a standard diode laser housing.
To access reference lines for long-term absolute stabilization, the laser is widely tunable using two microring resonators in Vernier configuration.
Active stabilization of the laser frequency to the acetylene P11 absorption line at 1531~nm provides a best stability of $2.0\cdot10^{-10}$.
This scheme enables the desired long-term stability, demonstrated by continuous frequency stabilization for over 5 days with a residual drift below 12~MHz.
For completeness, we also present short-term stabilization of our laser.
This is realized by locking to a fiber-based optical frequency discriminator (OFD), which provides a best relative frequency stability of $1.5\cdot10^{-12}$ over an averaging time of 0.5 ms.

\section{Hybrid-integrated diode laser}

\subsection{Laser design}

\begin{figure}[tb]
    \centering
    \includegraphics[width=0.58\linewidth]{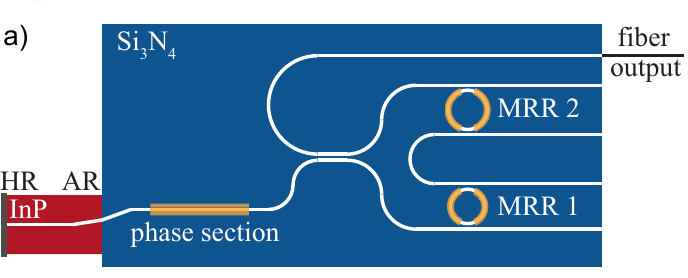}
    \includegraphics[width=0.4\linewidth]{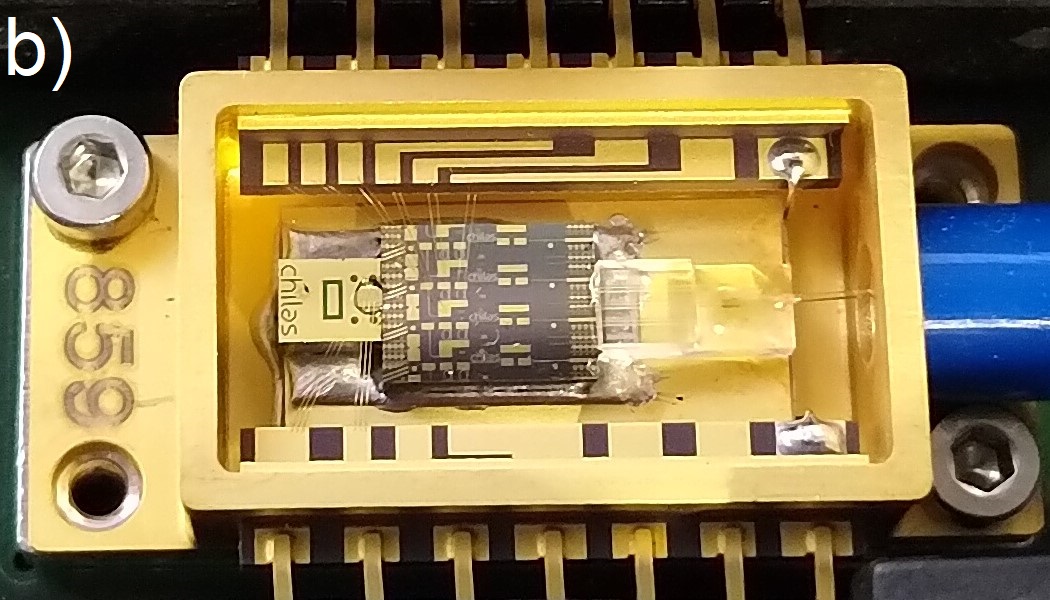}
    \caption{(a) Schematic view of the hybrid-integrated laser.
    The InP-based amplifier is coated with a high-reflection (HR) and an anti-reflection coating (AR). The Si$_3$N$_4$-based feedback chip contains two microring ring resonators (MRR).
    Heaters for thermal tuning are indicated in yellow.
    (b) Photograph of the hybrid laser assembly comprising the the amplifier, feedback chip and output fiber fixed together in a 14-pin butterfly package.}
    \label{fig:laser_schematic_photo}
\end{figure}

For the experiments, we use a narrow-linewidth hybrid-integrated tunable laser (Chilas CT3, similar to~\cite{VanRees_2020oe}) as schematically shown in Fig.~\ref{fig:laser_schematic_photo}(a).
The laser comprises a semiconductor chip, which is hybrid-integrated with a waveguide-based feedback circuit~\cite{fan_2016pj}.
The semiconductor chip contains an InP-based multi-quantum well optical amplifier, fabricated by the Fraunhofer Heinrich Hertz Institute.
The feedback circuit is based on low-loss Si$_3$N$_4$ waveguides with a TriPleX asymmetric double-stripe cross section~\cite{roeloffzen_2018jstqe}.
One mirror of the laser cavity is formed by a high-reflectivity (90\%) coating on the back facet of the optical amplifier.
The other mirror is a frequency-selective loop mirror, formed by two sequential microring resonators (MRRs) on the feedback circuit.
The circumferences of these resonators are 787~$\mu$m and 806~$\mu$m for MRR1 and MRR2, respectively.
According to the Vernier principle, this enables wavelength selection for single-frequency laser operation over 70.4~nm around the central wavelength of 1.55~$\mu$m.

In addition, these rings provide a low intrinsic linewidth by extending the effective cavity length with multiple passes through the MRRs.
The power coupling coefficient to the rings is designed as $\kappa^2 = 0.05$, which increases the optical effective length on resonance~\cite{liu_2001apl} to 23 and 24~mm for MRR1 and MRR2, respectively.
In total, the effective cavity roundtrip length is increased to approximately 7~cm when the cavity resonance and ring resonances are aligned.
To couple light out of the cavity to the output port, an 80\% directional coupler is part of the feedback circuit.
The output port is connected to a polarization-maintaining single-mode fiber, that is terminated with an FC/APC connector.

For thermal tuning, resistive heaters are located on top of the rings and a phase section.
The phase section enables tuning a cavity resonance to align with a common resonance of the MRRs.
All heaters can provide at least $2\pi$ phase shift by applying an electrical power of up to 1.0~W per heater.

The amplifier, feedback chip and output fiber are all fixed together on a common submount and placed on a thermoelectric cooler (TEC) element in a 14-pin butterfly package, shown in Fig.~\ref{fig:laser_schematic_photo}(b). 
The heaters and amplifier are wire bonded to separate pins.
For computer controlled heater tuning and low-noise laser operation, we use the internal heater driver of a Chilas tunable laser controller and external TEC and laser diode drivers.

\subsection{Laser characterization}
We characterized the laser in terms of output power and wavelength tuning.
For this purpose, we connected the laser via an optical isolator (Thorlabs IO-G-1550-APC) and a 90:10 fiber coupler to a photodiode power sensor (Thorlabs S144C) and an optical spectrum analyzer (OSA, ANDO AQ6317).

\begin{figure}[tb]
    \centering
    \includegraphics[width=0.49\linewidth]{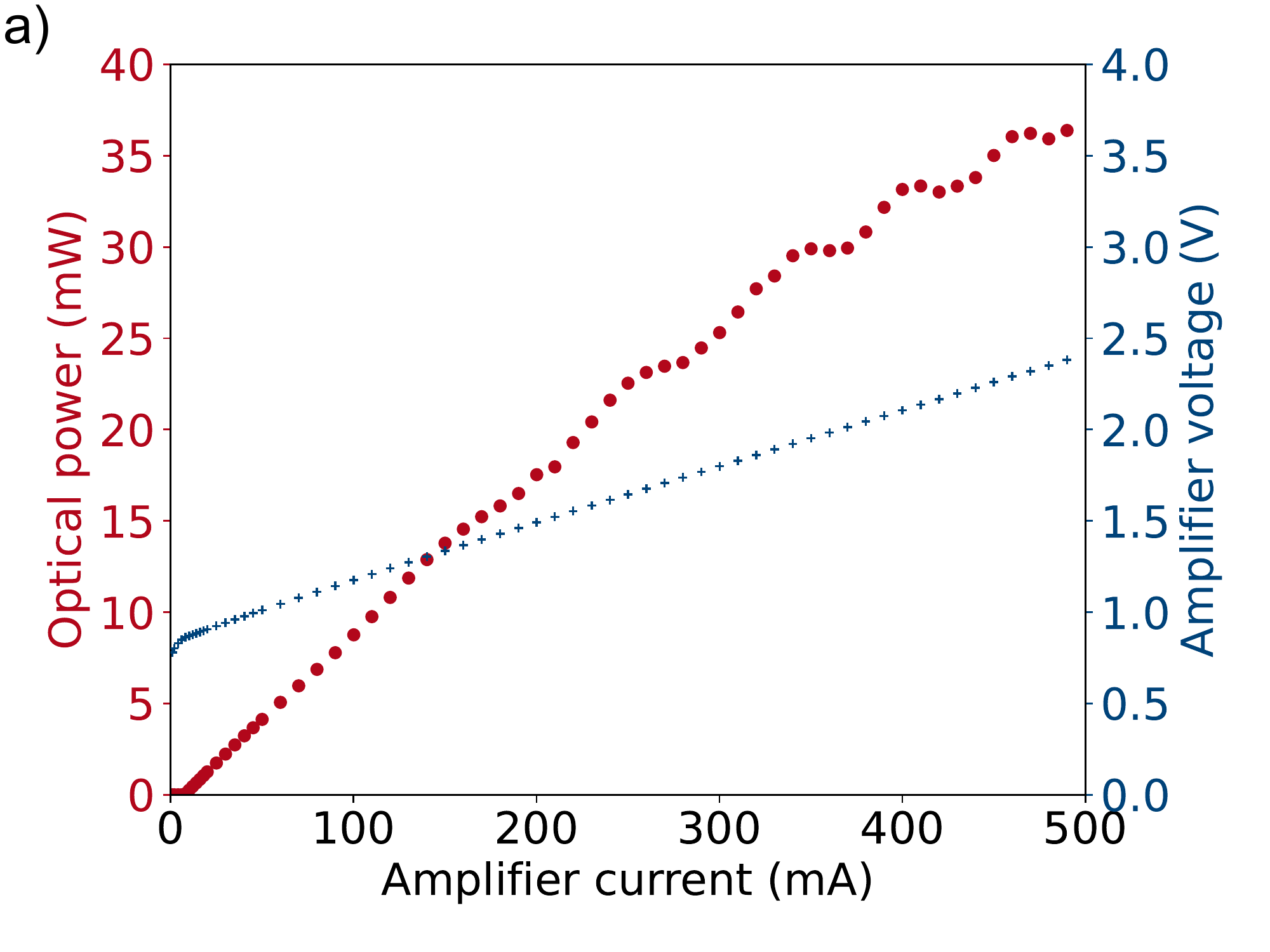}
    \includegraphics[width=0.49\linewidth]{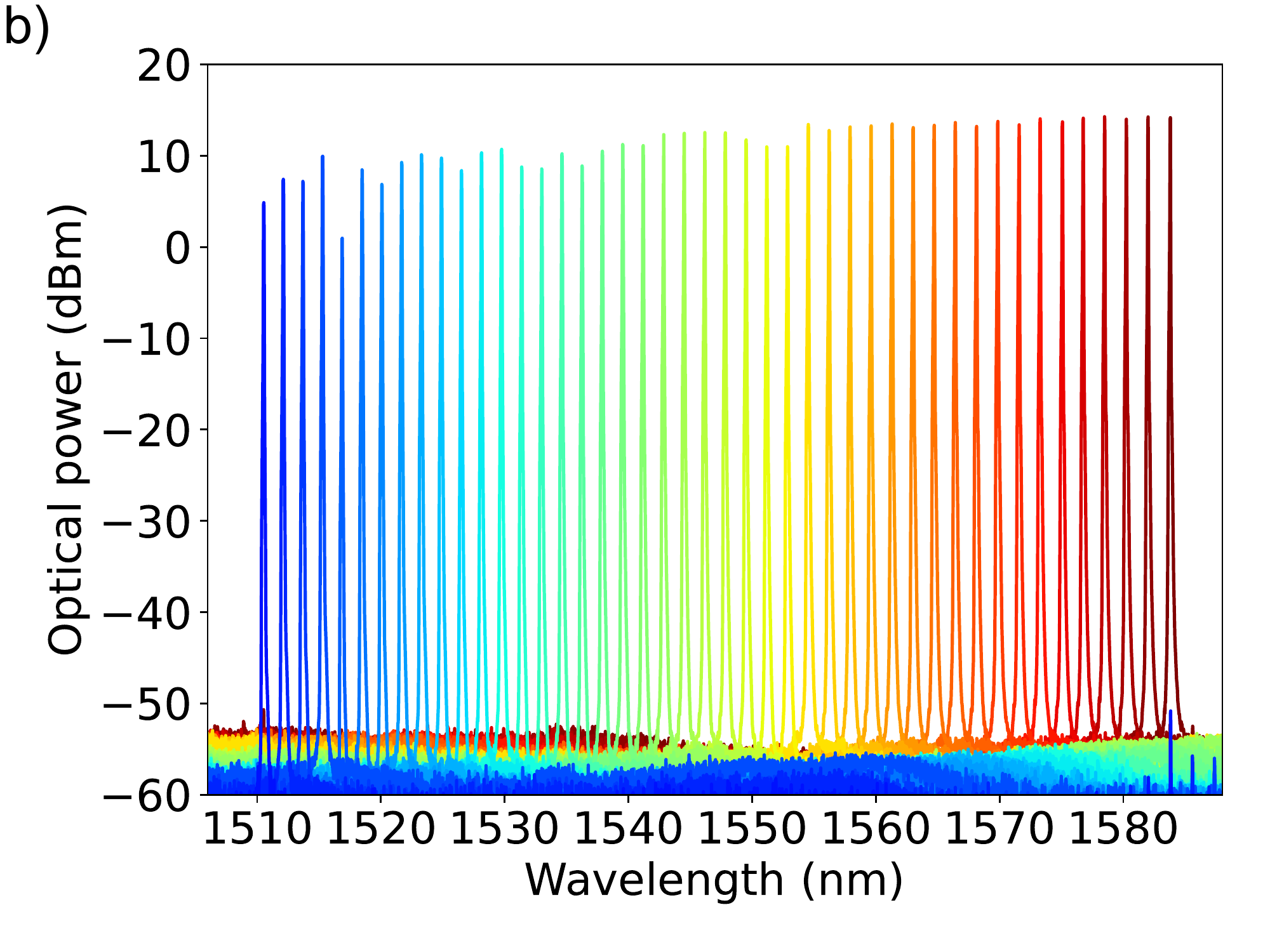}
    \caption{(a) Fiber-coupled output power (red dots) and amplifier voltage (blue crosses) versus amplifier current. The threshold current is 7.5~mA and the maximum measured fiber-coupled output power is 36.4~mW. (b) Superimposed laser spectra, as measured with an OSA set to 0.1~nm resolution bandwidth. The spectra are obtained by varying the heater power on MRR1 in steps of approximately 23~mW and then fine-tuning MRR1 and the phase section heater for maximum output power. The laser's wavelength coverage reaches from 1510.5 to 1583.8~nm. The side mode suppression ratio (SMSR) extracted from these spectra varies between 55 and 66~dB.}
    \label{fig:pi_curve_tuning}
\end{figure}

The fiber-coupled output power as function of the pump current and the corresponding amplifier voltages are plotted in Fig.~\ref{fig:pi_curve_tuning}(a).
The plotted output power is corrected for the measured transmission loss through the isolator and fiber coupler.
For exploring the maximum output power that can be generated with the laser, we provided a pump current of up to 500~mA using an external current source (ILX Lightwave LDX-3620).
For all other measurements we do not exceed the maximum specified current for the laser of 300~mA.
To obtain maximum output power and single-frequency operation, the phase section and rings were fine-tuned for each measurement.
The Vernier filter and the TEC element were set to 1569.9~nm and 20~\textdegree C, respectively.
Figure~\ref{fig:pi_curve_tuning}(a) indicates a threshold current of 7.5~mA and a maximum measured fiber-coupled output power of 36.4 mW.
Above threshold, the output power increases approximately linearly until power roll-off can be observed, which we address to local self-heating of the lasing region.

To characterize the wavelength tuning of the laser, we varied the heater power on ring resonator MRR1 and measured the laser spectra with the OSA set to 0.1~nm resolution bandwidth.
The diode pump current was set to the specified maximum of 300~mA for all measurements.
The measured power level of the OSA was calibrated using a separate power meter.
Figure~\ref{fig:pi_curve_tuning}(b) shows the superimposed spectra, when the heater power on MRR1 is changed in steps of approximately 23~mW and then fine-tuned together with the phase section for maximum output power.
As a result, the laser wavelength changes in steps of about 1.7~nm, which corresponds to the free spectral range of the other ring resonator (MRR2).
The wavelength coverage as shown in Fig.~\ref{fig:pi_curve_tuning}(b) spans from 1510.5 up to 1583.8~nm.
This range of 73.3~nm agrees well with the calculated Vernier tuning range.
The measured spectra display a high side mode suppression ratio (SMSR) between 61 and 66~dB throughout the tuning range of the laser, except for one outlier at the lower edge of the tuning range with a SMSR of 55 dB.
We have also verified single-mode operation using a high-resolution OSA (Finisar 1500S, spectral resolution 180~MHz).
The high SMSR confirms excellent single-mode operation over the full tuning range.

\section{Frequency stabilization}
For active frequency stabilization, laser frequency deviations from a set target have to be detected using a frequency discriminator.
Next, a feedback loop is required to set the laser frequency back to its target value.
To adjust the laser frequency, this laser offers multiple tuning mechanisms, which differ in tuning sensitivity and bandwidth.
In an earlier measurement using a similar laser, we found that the bandwidth for thermal tuning by the phase section is limited to approximately 50~kHz, while tuning the diode laser frequency by the amplifier current provides a higher bandwidth of at least 0.33~MHz, which was limited by the equipment used for that measurement~\cite{winkler2021frequency}.
To reduce noise up to MHz Fourier frequencies, we apply a feedback signal to the amplifier current.

We investigate two different fiber-coupled frequency discriminators that can be conveniently coupled to the laser's fiber output.
For stabilization on sub-second timescales, we use a commercial ultra-stable fiber-based frequency discriminator, which provides many closely spaced reference lines with a steep discriminator slope.
The availability of many reference lines, that are much closer spaced than the laser's cavity modes, enables locking for any Vernier heater setting.
An additional advantage of a fiber-based interferometer is that large path imbalances can be achieved to increase the frequency discriminator slope~\cite{Kefelian_ol2009}.
Maximizing this slope is important, since the highest achievable frequency stability is determined by the discriminator slope and its noise contribution, in the limit of large servo gain~\cite{Day_1992jqe}.
However, providing absolute stability is not possible, because long-term drift of the reference lines cannot be avoided completely, even with excellent thermal shielding.

To investigate long-term absolute frequency stabilization over multiple days, we use an acetylene absorption line as the second frequency reference.
In this case, locking requires tuning the laser wavelength to a single, selected absorption line.
The success of long-term stabilization critically depends on the absence of mode hops.
Therefore, we also characterize the frequency stability of the free-running laser, and demonstrate mode-hop free operation over multiple days.

\subsection{Sub-second frequency stabilization by locking to a fiber-based OFD} \label{short-term}

\begin{figure}[tb]
    \centering
    \includegraphics[width=0.55\linewidth]{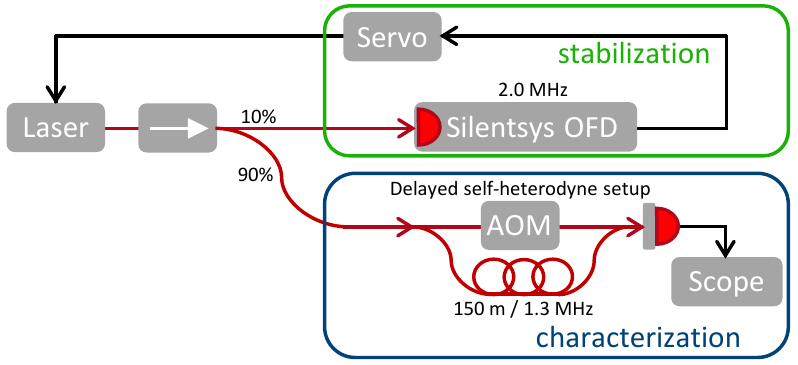}
    \caption{Frequency stabilization setup using a fiber-based optical frequency discriminator (OFD) and a setup for frequency noise characterization.
    The laser's output light first passes a fiber-optic isolator ($\rightarrow$) and a 90/10 fiber coupler, which directs 90\% of the input light to the characterization setup and 10\% to the stabilization setup.
    Here the OFD, which serves as a frequency reference for stabilization, generates an error signal, which is fed into a servo controller for feedback to the laser.
    This feedback signal is added to the laser drive current.
    For frequency noise characterization, a delayed self-heterodyne setup is used, which consists of an acousto-optic modulator (AOM) and a 150-m long fiber delay.
    The photodiode beat signal from this setup is recorded on an oscilloscope and processed to retrieve the frequency stability.}
    \label{fig:setup_ofd}
\end{figure}

The setup to stabilize the laser frequency on sub-second time scale, using a fiber-based optical frequency discriminator (Silentsys OFD), is shown in Fig.~\ref{fig:setup_ofd}.
To show typical stabilization behavior, the laser current and TEC were set to 100~mA and 25~\textdegree C, respectively for these measurements.
For optimum free-running operation, all heaters were turned off to remove any noise that might be added by the heater driver.
The fiber-coupled laser output first passes an optical isolator and then a 90/10 fiber-optic coupler.
90\% of the light is directed to a delayed self-heterodyne setup for analysis and 10\% is directed to the OFD for stabilization.
The OFD is an ultra-low noise optical frequency discriminator based on a fiber-interferometric process with a free spectral range of 2~MHz.
To improve its thermal stability, the OFD is placed in a thermally insulated enclosure with active temperature control.
The error signal from the OFD is fed to a servo controller equipped with double integrators (Liquid Instruments Moku:Lab, PID), which generates a feedback signal.
This feedback signal is added to the laser's amplifier current through the modulation input of the diode current driver (Koheron DRV200).

We characterize the frequency stability of the laser with out-of-loop delayed self-heterodyne measurements, using a similar setup as described in~\cite{Chauhan_2021nc}.
The setup comprises a fiber delay of 150~m and an acousto-optic modulator (AOM, Opto-Electronic MT80-IIR30-Fio-PM0.5-J1-A-Ic2) driven at 80~MHz.
The resulting beat signal is recorded using a balanced photodetector (Thorlabs PDB450C set to 150 MHz~bandwidth) connected to an oscilloscope (PicoScope 5444D) and processed to retrieve the frequency-noise power spectral density (PSD) of the laser.
To retrieve the PSD at high noise frequencies more accurately~\cite{Yuan_2022oe}, the PSD for multiple time segments is averaged, after dividing the original time trace in $2^n$ non-overlapping segments, where $n=0$ for the lowest Fourier frequencies up to $n=10$ at the highest frequencies.
Measurements of the PSD were carried out for the free running laser and for the locked laser.
When the laser was locked to the OFD, we optimized the settings of the servo controller to reduce the PSD in the frequency range below 100~kHz.

\begin{figure}[tb]
    \centering
    \includegraphics[width=0.49\linewidth]{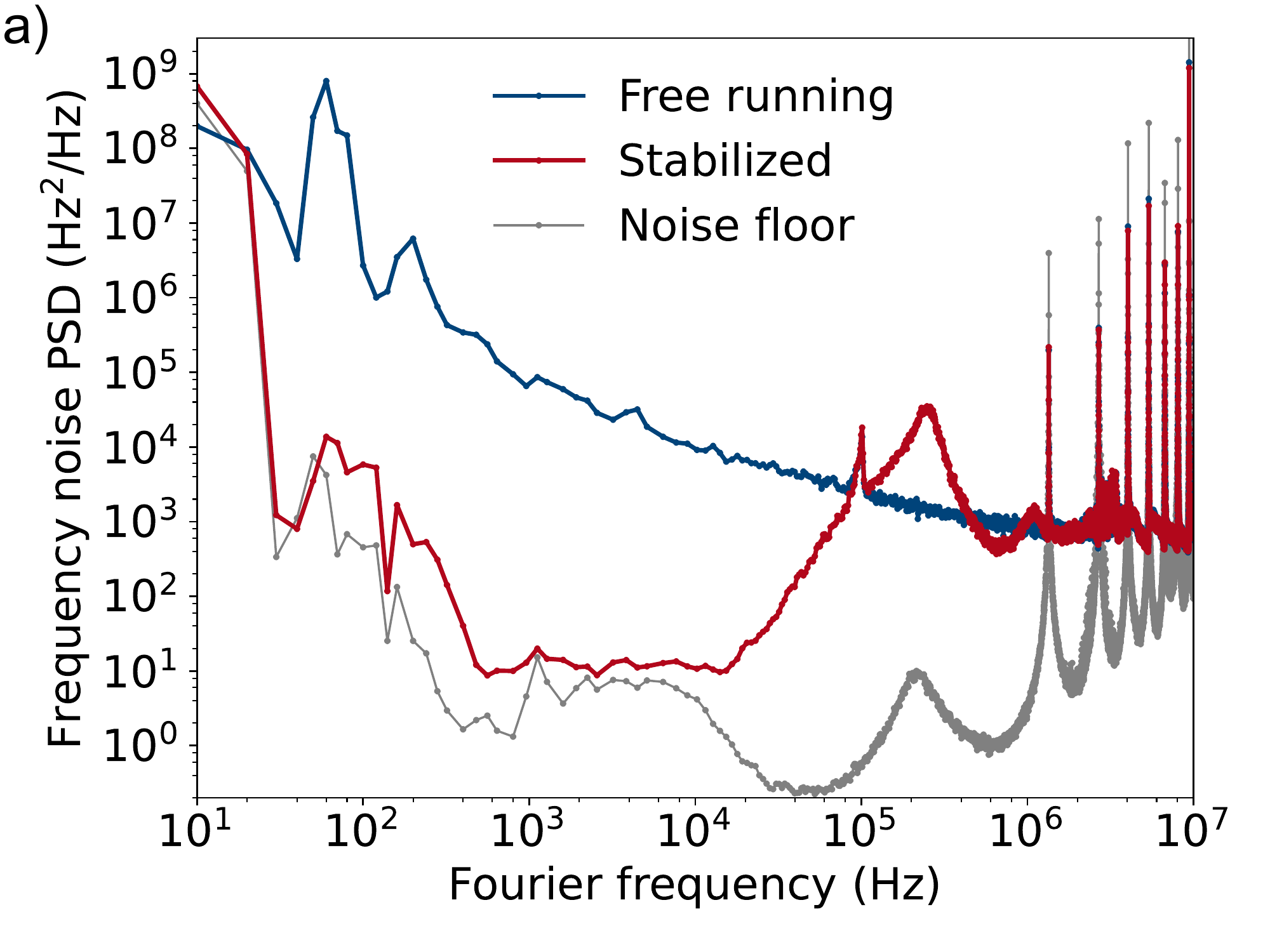}
    \includegraphics[width=0.49\linewidth]{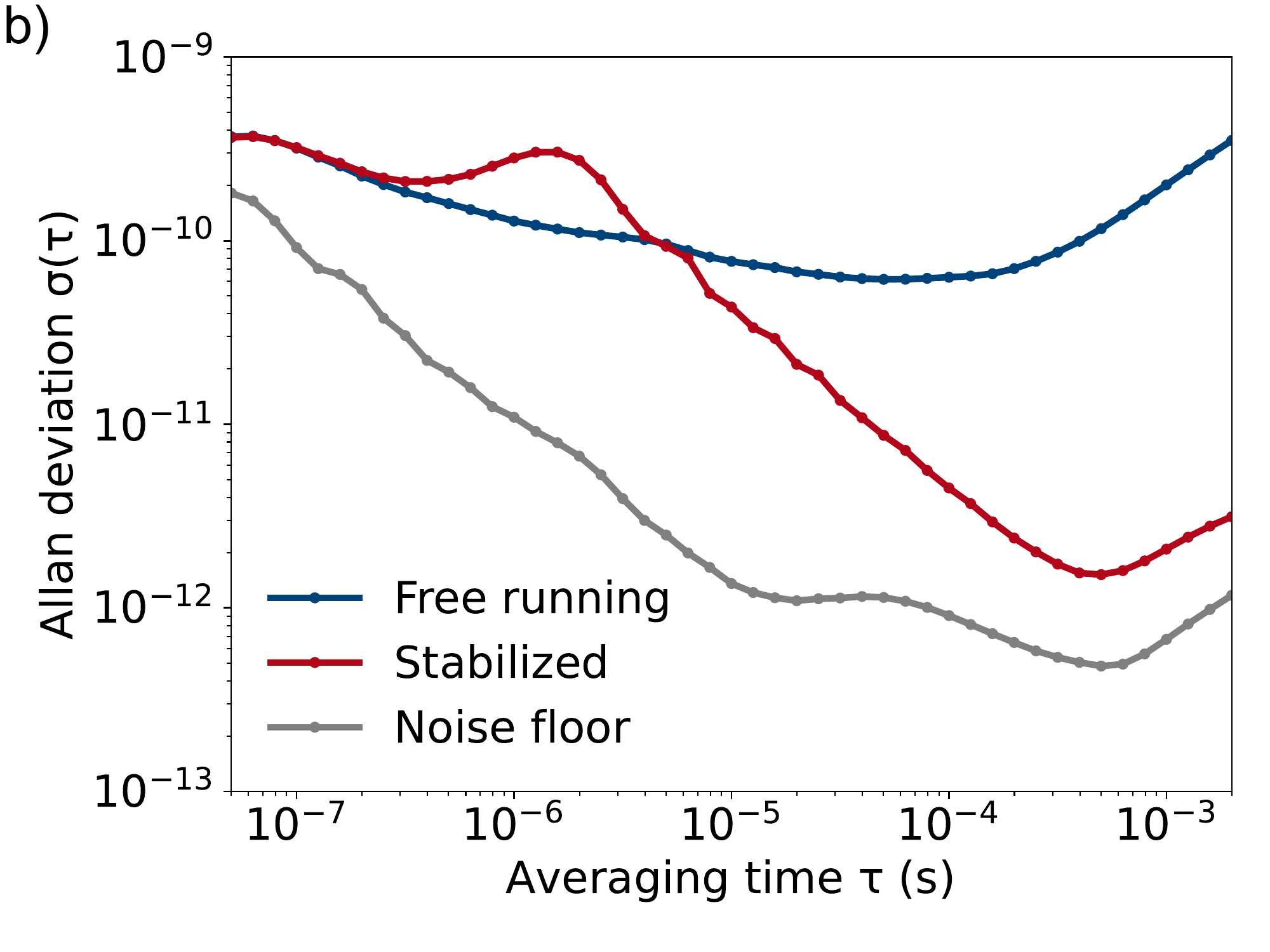}
    \caption{(a) The single-sided frequency-noise power spectral density (PSD) retrieved for the free running laser (blue trace) in comparison with the frequency stabilized laser (red trace) and the setup noise floor (gray trace).
    Frequency stabilization strongly reduces noise for Fourier frequencies up to $10^5$~Hz.
    Toward higher Fourier frequencies, the measured noise reaches approaches a white noise level of approximately 0.7~kHz$^2$/Hz, which corresponds to an intrinsic linewidth of 2~kHz.
    (b) Allan deviation for the free running laser (blue trace) in comparison with the frequency stabilized laser (red trace) and the noise floor of the system (gray trace).
    The Allan deviation of the free-running laser indicates a best stability of $6.2\cdot10^{-11}$ for an averaging time of 50~$\mu$s.
    Locking the laser frequency to the OFD improves the stability for averaging times $>7~\mu$s, reaching a minimum of $1.5\cdot10^{-12}$ over 0.5 ms.}
    \label{fig:FND_allan_shortterm}
\end{figure}

Figure~\ref{fig:FND_allan_shortterm}(a) shows the recorded frequency noise PSD of the free-running laser and the stabilized laser, as well as the noise floor of the measurement setup.
For the free-running laser, the PSD decreases with Fourier frequency, which is typical for thermo-refractive noise and noise stemming from thermal drifts, acoustic vibrations and flicker noise from electronics.
At high Fourier frequencies, beyond 2~MHz, the noise appears to reach a white noise level of 0.7~kHz$^2$/Hz.
Multiplying this level with $\pi$ yields an upper limit of 2~kHz for the intrinsic linewidth.
Noise occurring at a frequency of 1.3~MHz and its multiples cannot be resolved due to fringes, which is inherent to self-heterodyning with a 150~m fiber delay length.

Next, we measured the frequency noise when the laser frequency was stabilized to the OFD, also shown in Fig.~\ref{fig:FND_allan_shortterm}(a).
Stabilizing the laser frequency to the OFD strongly reduces the noise, especially for Fourier frequencies between 0.5 and 18~kHz, where an average noise level of 12~Hz\(^2\)/Hz is reached.
At Fourier frequencies around 250~KHz, a servo bump is clearly visible, which we address to the limited bandwidth of the control loop.

To investigate whether the measured frequency noise is limited by the characterization setup, we estimated the noise floor of the setup by removing the delay line to balance the path lengths in both arms of the delayed self-heterodyne setup.
Note, by removing the delay line, we may also have removed acoustic or thermal fluctuations in this long fiber from the noise floor measurement.
A small contribution of the laser noise is still present in the measured noise floor, which can be seen in Fig.~\ref{fig:FND_allan_shortterm}(a) by the servo bump at 250~kHz, indicating that the path lengths were not exactly balanced.
From this estimation of the noise floor, we find that the noise floor is sufficiently low for noise measurements at Fourier frequencies between 60~Hz and 1~kHz, and above 10~kHz.
Below 60~Hz, the noise floor increases steeply, probably due to pick-up of acoustic and thermal noise in the delayed self-heterodyne setup.
Between Fourier frequencies of 1 and 10 kHz, the measured noise PSD of 13~Hz$^2$/Hz is also close to the noise floor of the measurement setup.
To investigate whether the noise PSD was here limited by the oscilloscope, we used a different oscilloscope (Rohde \& Schwarz RTE1024) and indeed found a slightly lower noise PSD of 5~Hz$^2$/Hz, while the noise floor in this frequency range was reduced to about 0.2~Hz$^2$/Hz.

To display the frequency stability for different averaging times, Fig.~\ref{fig:FND_allan_shortterm}(b) shows the Allan deviation for the free-running and the stabilized laser.
Both traces are calculated from the recorded PSD, using Eq.~5.12 from~\cite{rutman1978characterization}.
The Allan deviation is calculated using only the data starting from 30~Hz, as the recorded PSD for the stabilized laser below that frequency is limited by the noise floor of the delayed self-heterodyne setup.
The fringes at 1.3~MHz and its multiples were also excluded as they are an artifact of the measurement method.
For the free-running laser, the Allan deviation reaches a minimum of $6.2\cdot10^{-11}$ for an averaging time of 50 $\mu$s.
Stabilization to the OFD improves the laser's frequency stability by up to two orders of magnitude, reaching a best value of $1.5\cdot10^{-12}$ for an averaging time of 500~$\mu$s.

\subsection{Long-term absolute frequency stabilization by locking to acetylene}
For all applications using lasers for measurements, stability over the entire duration of the measurement is required.
Since many applications require long-term absolute stability, we also investigate the frequency stability over longer times, up to several days.
For this purpose, we require an absolute frequency reference within the laser's spectral coverage.
A suitable reference is an acetylene gas cell, as the influence of changes in environmental conditions, such as temperature, pressure and electromagnetic fields, on the absorption lines is negligible~\cite{Swann_2000josab} compared to the measurement accuracy here.

\begin{figure}[tb]
    \centering
    \includegraphics[width=0.55\linewidth]{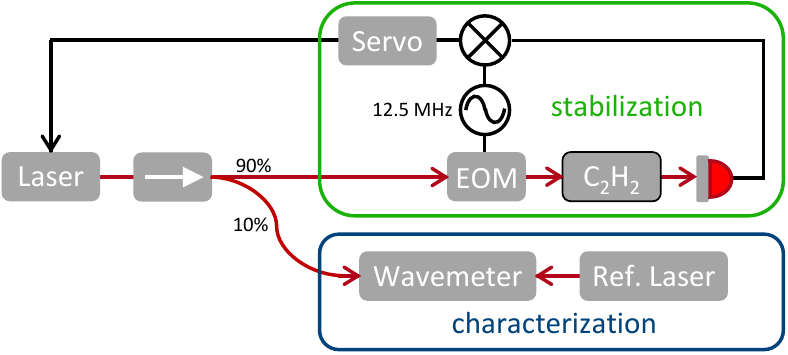}
    \caption{Frequency stabilization setup using a fiber-coupled acetylene (C$_2$H$_2$) gas cell. 
    Behind the isolator ($\rightarrow$), the laser light passes through a 90/10 fiber coupler, which directs 90\% of the input light to the stabilization setup.
    This setup for top-of-fringe locking comprises an electro-optic modulator (EOM), a fiber-coupled acetylene ($^{12}$C$_2$H$_2$) gas cell, a fast photodiode and an electronic servo controller.
    The controller generates a 12.5~MHz signal for modulating the light and for demodulating the photodiode signal.
    Feedback from the servo controller is added to the laser's amplifier current.
    To characterize the frequency stability of the laser, 10\% of the output light is directed to a wavelength meter, which is hourly calibrated using a reference diode laser stabilized to a Doppler-free rubidium line.}
    \label{fig:setup_acetylene}
\end{figure}

A central prerequisite for successful long-term stabilization is that the laser emission does not hop to other cavity modes, as mode hops are hard to correct for in any stabilization scheme.
Even a mode hop to a neighboring mode would shift the laser frequency outside the capture range of the frequency reference, since the calculated 4.6~GHz minimum mode distance for this laser is larger than the sub-GHz linewidth of an acetylene reference line.
To avoid mode hops, a high passive stability of the laser cavity length is required.
Hybrid integration of the diode amplifier with the feedback chip aims on providing this stability with permanent and robust fixation of the aligned waveguide circuits.
To detect any mode hops and, to more detail, record the laser's emission frequency over several days, we use a broadband and ultra-precise wavelength meter.

Fig.~\ref{fig:setup_acetylene} shows the setup used for long-term stabilization.
The laser current and TEC were again set to 100~mA and 25~\textdegree C, respectively.
The fiber-coupled laser output first passes an optical isolator and then a 90/10 fiber-optic coupler.
To maximize the photodetector signal, the largest fraction of the laser output is used for frequency stabilization.
The stabilization scheme is based on top-of-fringe locking using the principle of frequency modulation (FM) spectroscopy~\cite{Bjorklund_1983apb}.
The light passes an electro-optic phase modulator (EOM, Thorlabs LN65-10-P-A-A-BNL) driven at a modulation frequency of 12.5~MHz.
The modulated light then propagates through a fiber-coupled gas cell (Wavelength References C2H2-12-H(16.5)-4-FCAPC), filled with acetylene ($^{12}$C$_2$H$_2$), which serves as the frequency reference.
Although a Doppler-free measurement of a saturated acetylene absorption line would provide the narrowest reference~\cite{deLabachelerie_1995ol}, this also requires a gas pressure that is lower than available for standard fiber-coupled reference cells.
We used a fiber-coupled gas cell with a path length of 16.5~cm and a pressure of 4~Torr, that provides a narrow line of approximately 600~MHz width and 74\% peak absorption, as predicted with a HITRAN-based simulation~\cite{goldenstein_2017jqsrt}.
The power transmitted through the gas cell is measured using an amplified photodetector (Thorlabs PDB450C).
The detector signal is fed into a locking module (Toptica DigiLock 110), which demodulates the signal and provides feedback to the laser current using a proportional integral differential (PID) controller.

To measure the long-term stability of the laser, 10\% of the laser output is directed to a Fitzeau-interferometer based wavelength meter (HighFinesse {\AA}ngstrom WS Ultimate 30 IR).
The specified relative accuracy is $3\cdot10^{-7}$ and the absolute accuracy is 30~MHz, however, we suspect that the measurement accuracy is well below this specification~\cite{Saleh_2015ao, konig2020performance}.
The sampling time of the wavelength meter was set automatically with typical sampling times between 50 and 200~ms.
To counteract the long-term drift of the wavelength meter, it was calibrated every hour using a reference diode laser stabilized to a Doppler-free rubidium line ($^{85}$Rb, $D_2$, crossover $F=3\rightarrow{} F'=3, 4$) at 780.2~nm.
For increasing the accuracy also between calibrations, all measured data points are corrected by assuming a linear drift of the wavelength meter between these hourly calibration points.

\begin{figure}[tb]
    \centering
    \includegraphics[width=0.49\linewidth]{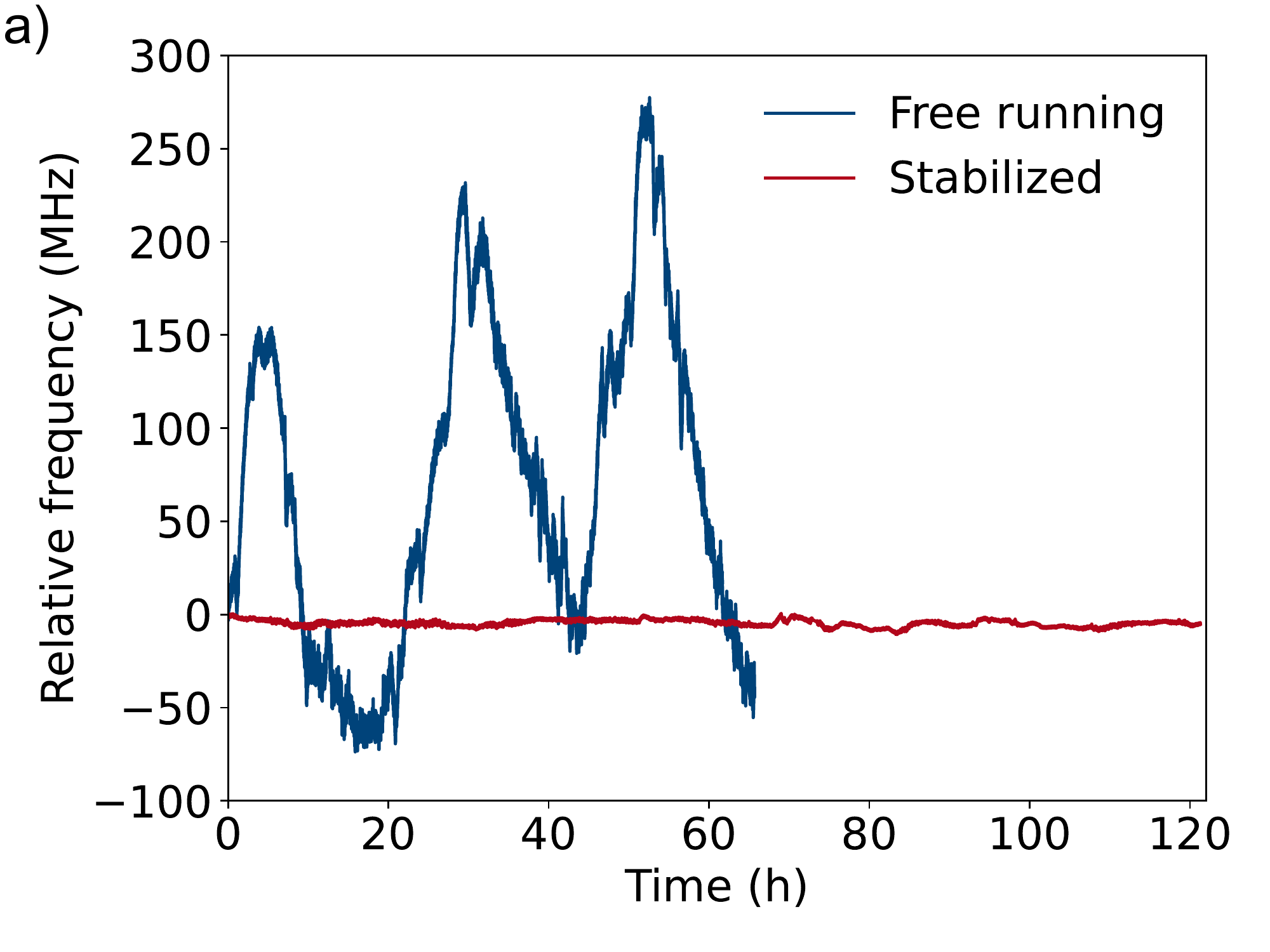}
    \includegraphics[width=0.49\linewidth]{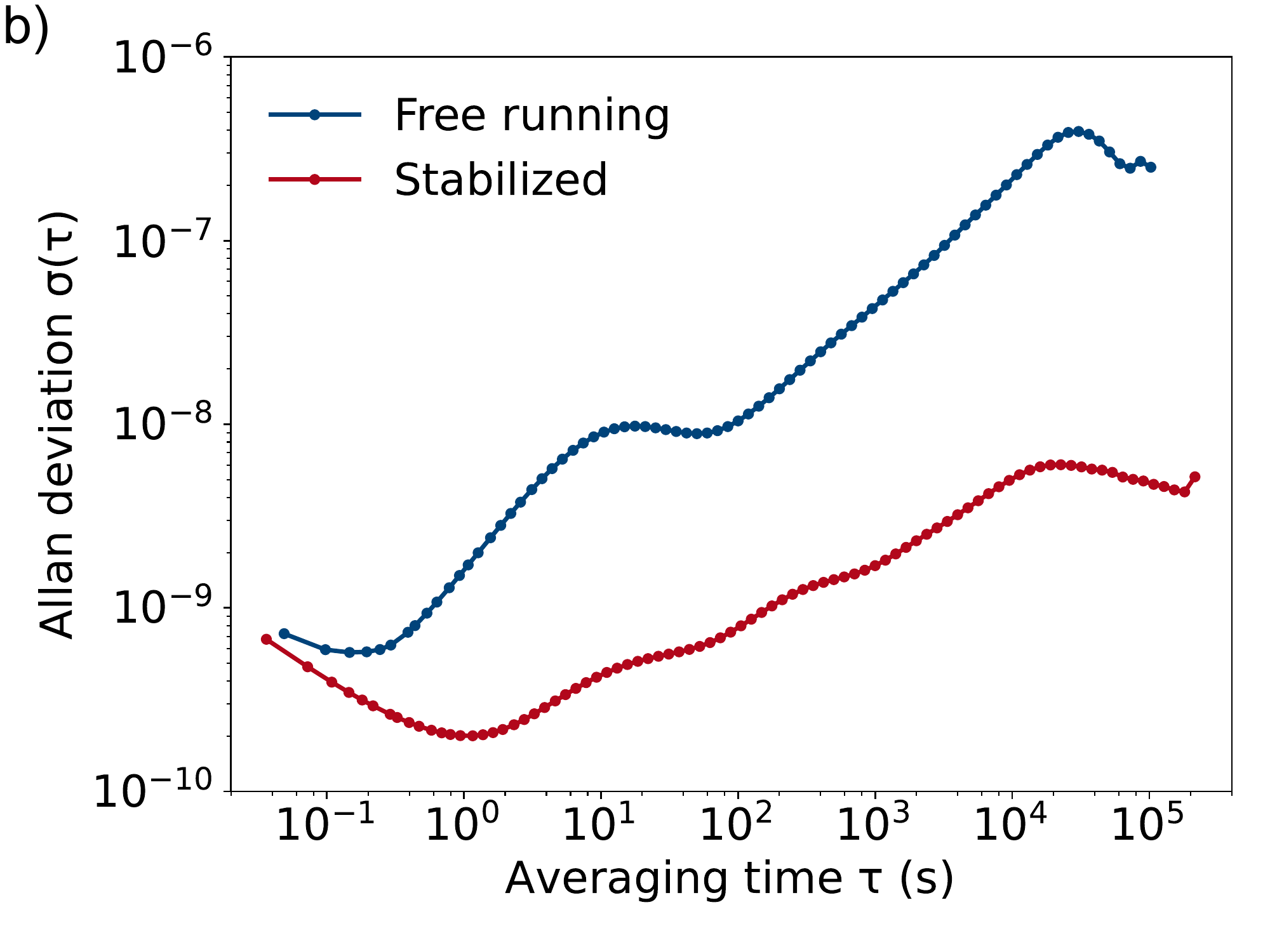}
    \caption{(a) Recorded emission frequency, relative to the initial value, of the free-running laser over 65 hours (blue trace) and of the stabilized laser over 121 hours (red trace).
    The frequency of the free-running laser does not show any discontinuities, which confirms the absence of mode hops. The frequency drift remains within a range of 352~MHz, which is small compared to the free spectral range of the laser cavity.
    The frequency variations follow the 24-hour rhythm of the room temperature.
    Frequency stabilization to an acetylene absorption line greatly reduces the drift to a range of less than 12~MHz.
    (b) The overlapping Allan deviation of the free-running laser, as shown with the blue trace, indicates a best frequency stability of $5.7\cdot10^{-10}$ for an averaging time of 0.15 seconds.
    Locking the laser frequency to the acetylene absorption line, as shown in red, improves the long-term stability, reaching a minimum of $2.0\cdot10^{-10}$ over 1.2~seconds.}
    \label{fig:frequency_allan_longterm}
\end{figure}

To characterize the frequency stability of the free-running laser, we first tuned the laser wavelength to 1531.5879~nm by setting the heaters on the MRRs and the phase section accordingly.
This wavelength is at the P11 absorption line, to which we later locked the laser.
Figure~\ref{fig:frequency_allan_longterm}(a) shows, as the blue trace, the recorded frequency deviations of the free-running laser over a measurement time of 66~hours.
We note that, if any mode hop had occurred during that time, the laser's frequency would deviate by a discontinuous step at least ten times the 400~MHz vertical range chosen for this plot.
However, the trace remaining continuously within the displayed vertical range proves that no mode hop occurred for the entire duration of the measurement.
This high passive stability is obtained by the robust hybrid integration of the diode amplifier with the feedback circuit and makes this laser well suited for long-term frequency stabilization.

In free-running operation, the laser frequency drifts within a frequency range of 352~MHz over three days.
We found that the frequency drift is strongly correlated with the variation in lab temperature of 1.9~K over 24-hour intervals.
This is expected, as the waveguide effective index and therefore the effective length of the laser cavity, are temperature dependent.
Although the temperature of the laser submount is stabilized by a TEC controller, the local temperatures in the waveguide circuit can vary due to lab temperature variations.
The observed frequency drift is of the same order of magnitude as described in~\cite{GonzalezGuerrero_2022jlt} for a similar laser with several different heater settings, which drifted no more than 120~MHz in 90~minutes.

To counteract the drift, we lock the laser's emission frequency to the center of the acetylene absorption line by providing feedback to the diode current.
The recorded frequency deviations for the stabilized laser are shown in red in Fig.~\ref{fig:frequency_allan_longterm}(a) for a recording time of over 5 days.
It can be seen that the laser remains locked to the absorption line over the 121-hour duration of the measurement.
The frequency deviations are within a range of 12~MHz, which is a great reduction of more than one order of magnitude compared to the free running laser.
The residual variations in frequency displayed by the wavelength meter can be partly addressed to the thermal sensitivity of the wavelength meter~\cite{Saleh_2015ao}, because we observe a remaining correlation with room temperature.

To further analyze the frequency stability on different time scales, Fig.~\ref{fig:frequency_allan_longterm}(b) displays the overlapping Allan deviation of the frequency deviations of the free-running and the stabilized laser.
On short timescales below 0.1~s, we find that the measured stability is only slightly improved.
We suspect the limit here is given by the wavelength meter, since~\cite{couturier2018laser} reports a measurement of an ultra-stable laser using a higher resolution variant of the wavelength meter used here, where only a slightly lower level of stability is measured.
Over longer averaging times, we observe that the frequency of the locked laser is more stable than for the free-running laser, since drifts over longer times can be very well detected and corrected with this locking scheme.
The best frequency stability for the free-running laser is $5.7\cdot10^{-10}$ over an averaging time of 0.15 seconds.
For the locked laser, the frequency stability is improved on all timescales, reaching a minimum of $2.0\cdot10^{-10}$ over an averaging time of 1.2~seconds.

\section{Conclusion and outlook}

We have demonstrated the frequency stabilization of a widely-tunable hybrid-integrated diode laser, both on short and long time scales.
To provide a steep error signal for short-term stability, we used a fiber-based optical frequency discriminator (OFD) as a reference.
By locking the laser to the OFD, the laser's fractional frequency stability reaches a relative stability of $1.5\cdot10^{-12}$ over an averaging time of 0.5~ms.
For absolute long-term stability, we locked the laser frequency to an acetylene absorption line, which limits the frequency deviations of the laser to a range of less than 12~MHz over 5~days.
To further improve the frequency stability, one could use a Doppler-free absorption line~\cite{deLabachelerie_1995ol} which provides a steeper error signal than the pressure-broadened acetylene line used here.
Similarly to the approach described in~\cite{Sakai_1991ptl}, the two references could also be combined by stabilizing the laser to the OFD which in turn could be stabilized to an acetylene absorption line to avoid long-term drifts.
As a next step, a fully integrated laser system with reduced frequency noise or absolute stability could be realized by combining a hybrid-integrated laser with an on-chip reference, for example using a spiral resonator~\cite{Lee_2013nc,Liu_2021optica}, a thermally-insensitive ring-resonator readout~\cite{Zhao_2021opt} or a full spectroscopic unit~\cite{Hummon_2018opt,Zektzer_2020lpr}.

Long-term stable, low-noise lasers in the telecom range are of great importance for data transmission, where the carrier frequency must comply to a predefined grid~\cite{GonzalezGuerrero_2022jlt}. 
Such chip-integrated lasers will also be of great interest for precision metrology, optical clocks and quantum technology.
Typically, these applications require tuning and stabilizing a laser’s emission frequency to address particular atomic or molecular transitions, many of them in the visible spectral region.
Recently, the wavelength coverage of widely-tunable chip-integrated lasers has been greatly expanded.
Heterogeneous integration of III-V materials with Si$_3$N$_4$ waveguides has increased its wavelength coverage from the telecom wavelength range toward the 1.0 µm range~\cite{Bovington_2014oe, Tran_2022nat}.
On the other hand, hybrid integration with Si$_3$N$_4$ and Al$_2$O$_3$ waveguides poses less restrictions on material properties and fabrication technologies, which has enabled hybrid integrated ECDLs to enter also the visible spectral range~\cite{Franken_2021ol, Winkler_2023spie, Franken_2023arxiv}.
Such lasers, stabilized to according references, e.g., to iodine, strontium and rubidium lines, will open up new possibilities for advanced photonics circuits in terms of portability, scalability and stability.

\section*{Funding}
This work is in part supported by the European Union’s Horizon 2020 research and innovation program under Grant 780502 (3PEAT) and in part by the Netherlands Enterprise Agency under Grant PPS\textunderscore2020\textunderscore90 (Unlocking hybrid photonics for ultra-precise laser applications).

\section*{Acknowledgments}
We thank Manfred Hager for helpful discussions and Thomas Puppe for providing the wavelength meter.

\bibliographystyle{IEEEtran}

\begin{thebibliography}{1}
\bibitem{Niffenegger_2020nat}
\href{https://doi.org/10.1038/s41586-020-2811-x}{R.~J. Niffenegger, J.~Stuart, C.~Sorace-Agaskar, D.~Kharas, S.~Bramhavar, C.~D. Bruzewicz, W.~Loh, R.~T. Maxson, R.~McConnell, D.~Reens, G.~N. West, J.~M. Sage, and J.~Chiaverini, ``{Integrated multi-wavelength control of an ion qubit},'' \emph{Nature}, vol. 586, no. 7830, pp. 538--542, 2020.}

\bibitem{Mehta_2020nat}
\href{https://doi.org/10.1038/s41586-020-2823-6}{K.~K. Mehta, C.~Zhang, M.~Malinowski, T.-L. Nguyen, M.~Stadler, and J.~P. Home,``{Integrated optical multi-ion quantum logic},'' \emph{Nature}, vol. 586, no. 7830, pp. 533--537, 2020.}

\bibitem{Mahmudlu_2023np}
\href{https://doi.org/10.1038/s41566-023-01193-1}{H.~Mahmudlu, R.~Johanning, A.~van Rees, A.~K. Kashi, J.~P. Epping, R.~Haldar, K.-J. Boller, and M.~Kues, ``Fully on-chip photonic turnkey quantum source for entangled qubit/qudit state generation,'' \emph{Nature Photonics}, 2023.}

\bibitem{newman_2019o}
\href{https://doi.org/10.1364/OPTICA.6.000680}{Z.~L. Newman, V.~Maurice, T.~Drake, J.~R. Stone, T.~C. Briles, D.~T. Spencer, C.~Fredrick, Q.~Li, D.~Westly, B.~R. Ilic, B.~Shen, M.-G. Suh, K.~Y. Yang, C.~Johnson, D.~M.~S. Johnson, L.~Hollberg, K.~J. Vahala, K.~Srinivasan, S.~A. Diddams, J.~Kitching, S.~B. Papp, and M.~T. Hummon, ``Architecture for the photonic integration of an optical atomic clock,'' \emph{Optica}, vol.~6, no.~5, pp. 680--685, 2019.}

\bibitem{marpaung_2019np}
\href{https://doi.org/10.1038/s41566-018-0310-5}{D.~Marpaung and J.~Yao, ``Integrated microwave photonics,'' \emph{Nature Photonics}, vol.~13, pp. 80--90, 2019.}

\bibitem{Botter_2022sciadv}
\href{https://doi.org/10.1126/sciadv.abq2196}{R.~Botter, K.~Ye, Y.~Klaver, R.~Suryadharma, O.~Daulay, G.~Liu, J.~van~den Hoogen, L.~Kanger, P.~van~der Slot, E.~Klein, M.~Hoekman, C.~Roeloffzen, Y.~Liu, and D.~Marpaung, ``{Guided-acoustic stimulated Brillouin scattering in silicon nitride photonic circuits},'' \emph{Science Advances}, vol.~8, eabq2196, 2022.}

\bibitem{lin_2018pj}
\href{https://doi.org/10.1109/JPHOT.2018.2842026}{Y.~Lin, C.~Browning, R.~B. Timens, D.~H. Geuzebroek, C.~G.~H. Roeloffzen, M.~Hoekman, D.~Geskus, R.~M. Oldenbeuving, R.~G. Heideman, Y.~Fan, K.-J. Boller, and L.~P. Barry, ``{Characterization of Hybrid InP-TriPleX Photonic Integrated Tunable Lasers Based on Silicon Nitride (Si$_3$N$_4$/SiO$_2$) Microring Resonators for Optical Coherent System},'' \emph{IEEE Photonics Journal}, vol.~10, no.~3, {1400108}, 2018.}

\bibitem{Zou_2020ol}
\href{https://doi.org/10.1364/OL.383137}{K.~Zou, Z.~Zhang, P.~Liao, H.~Wang, Y.~Cao, A.~Almaiman, A.~Fallahpour, F.~Alishahi, N.~Satyan, G.~Rakuljic, M.~Tur, A.~Yariv, and A.~E. Willner, ``{Higher-order QAM data transmission using a high-coherence hybrid Si/III–V semiconductor laser},'' \emph{Optics Letters}, vol.~45, no.~6, pp. 1499--1502, 2020.}

\bibitem{Dass_2022jlt}
\href{https://doi.org/10.1109/JLT.2022.3169446}{D.~Dass, A.~Delmade, L.~Barry, C.~G.~H. Roeloffzen, D.~Geuzebroek, and C.~Browning, ``{Wavelength \& mm-Wave Flexible Converged Optical Fronthaul With a Low Noise Si-Based Integrated Dual Laser Source},'' \emph{Journal of Lightwave Technology}, vol.~40, pp. 3307--3315, 2022.}

\bibitem{Tran_2019apl}
\href{https://doi.org/10.1063/1.5124254}{M.~A. Tran, D.~Huang, and J.~E. Bowers, ``{Tutorial on narrow linewidth tunable semiconductor lasers using Si/III-V heterogeneous integration},'' \emph{APL Photonics}, vol.~4, 111101, 2019.}

\bibitem{Tran_2020jstqe}
\href{https://doi.org/10.1109/JSTQE.2019.2935274}{M.~A. Tran, D.~Huang, J.~Guo, T.~Komljenovic, P.~A. Morton, and J.~E. Bowers, ``{Ring-Resonator Based Widely-Tunable Narrow-Linewidth Si/InP Integrated Lasers},'' \emph{IEEE Journal of Selected Topics in Quantum Electronics}, vol.~26, no.~2, 1500514, 2020.}

\bibitem{boller_2019Phot}
\href{https://doi.org/10.3390/photonics7010004}{K.-J. Boller, A.~van Rees, Y.~Fan, J.~Mak, R.~E.~M. Lammerink, C.~A.~A. Franken, P.~J.~M. van~der Slot, D.~A.~I. Marpaung, C.~Fallnich, J.~P. Epping, R.~M. Oldenbeuving, D.~Geskus, R.~Dekker, I.~Visscher, R.~Grootjans, C.~G.~H. Roeloffzen, M.~Hoekman, E.~J. Klein, A.~Leinse, and R.~G. Heideman, ``Hybrid integrated semiconductor lasers with silicon nitride feedback circuits,'' \emph{Photonics}, vol.~7, no.~1, 4, 2019.}

\bibitem{Fan_2020oe}
\href{https://doi.org/10.1364/OE.398906}{Y.~Fan, A.~van Rees, P.~J.~M. van~der Slot, J.~Mak, R.~M. Oldenbeuving, M.~Hoekman, D.~Geskus, C.~G.~H. Roeloffzen, and K.-J. Boller, ``{Hybrid integrated InP-Si$_3$N$_4$ diode laser with a 40-Hz intrinsic linewidth},'' \emph{Optics Express}, vol.~28, no.~15, pp. 21\,713--21\,727, 2020.}

\bibitem{Jin_2021np}
\href{https://doi.org/10.1038/s41566-021-00761-7}{W.~Jin, Q.-F. Yang, L.~Chang, B.~Shen, H.~Wang, M.~A. Leal, L.~Wu, M.~Gao, A.~Feshali, M.~Paniccia, K.~J. Vahala, and J.~E. Bowers, ``{Hertz-linewidth semiconductor lasers using CMOS-ready ultra-high-Q microresonators},'' \emph{Nature Photonics}, vol.~15, no.~5, pp. 346--353, 2021.}

\bibitem{Li_2021ol}
\href{https://doi.org/10.1364/OL.439720}{B.~Li, W.~Jin, L.~Wu, L.~Chang, H.~Wang, B.~Shen, Z.~Yuan, A.~Feshali, M.~Paniccia, K.~J. Vahala, and J.~E. Bowers, ``{Reaching fiber-laser coherence in integrated photonics},'' \emph{Optics Letters}, vol.~46, no.~20, pp. 5201--5204, 2021.}

\bibitem{Siddharth_2022apl}
\href{https://doi.org/10.1063/5.0081660}{A.~Siddharth, T.~Wunderer, G.~Lihachev, A.~S. Voloshin, C.~Haller, R.~N. Wang, M.~Teepe, Z.~Yang, J.~Liu, J.~Riemensberger, N.~Grandjean, N.~Johnson, and T.~J. Kippenberg, ``Near ultraviolet photonic integrated lasers based on silicon nitride,'' \emph{APL Photonics}, vol.~7, 046108, 2022.}

\bibitem{Corato_2023np}
\href{https://doi.org/10.1038/s41566-022-01120-w}{M.~Corato-Zanarella, A.~Gil-Molina, X.~Ji, M.~C. Shin, A.~Mohanty, and M.~Lipson, ``Widely tunable and narrow-linewidth chip-scale lasers from near-ultraviolet to near-infrared wavelengths,'' \emph{Nature Photonics}, vol.~17, pp. 157--164, 2023.}

\bibitem{ciddor1983two}
\href{https://doi.org/10.1088/0022-3735/16/12/024}{P.~Ciddor and R.~Duffy, ``Two-mode frequency-stabilised He-Ne (633 nm) lasers: studies of short- and long-term stability,'' \emph{Journal of Physics E: Scientific Instruments}, vol.~16, no.~12, pp. 1223--1227, 1983.}

\bibitem{rowley1990performance}
\href{https://doi.org/10.1088/0957-0233/1/4/006}{W.~Rowley, ``The performance of a longitudinal Zeeman-stabilised He-Ne laser (633 nm) with thermal modulation and control,'' \emph{Measurement Science and Technology}, vol.~1, no.~4, pp. 348--351, 1990.}

\bibitem{arnold1998simple}
\href{https://doi.org/10.1063/1.1148756}{A.~Arnold, J.~Wilson, and M.~Boshier, ``A simple extended-cavity diode laser,'' \emph{Review of Scientific Instruments}, vol.~69, no.~3, pp. 1236--1239, 1998.}

\bibitem{martin2016external}
\href{https://doi.org/10.1007/s00340-016-6575-9}{A.~Martin, P.~Baus, and G.~Birkl, ``External cavity diode laser setup with two interference filters,'' \emph{Applied Physics B}, vol. 122, 298, 2016.}

\bibitem{Krause_2020ao}
\href{https://doi.org/10.1364/AO.409308}{F.~Krause, E.~Benkler, C.~Nölleke, P.~Leisching, and U.~Sterr, ``Simple and compact diode laser system stabilized to Doppler-broadened iodine lines at 633 nm,'' \emph{Applied Optics}, vol.~59, pp. 10\,808--10\,812, 2020.}

\bibitem{Herbers_2022ol}
\href{https://doi.org/10.1364/OL.470984}{S.~Herbers, S.~Häfner, S.~Dörscher, T.~Lücke, U.~Sterr, and C.~Lisdat, ``Transportable clock laser system with an instability of $1.6\times10^{-16}$,'' \emph{Optics Letters}, vol.~47, pp. 5441--5444, 2022.}

\bibitem{Matei_2017prl}
\href{https://doi.org/10.1103/PhysRevLett.118.263202}{D.~G. Matei, T.~Legero, S.~Häfner, C.~Grebing, R.~Weyrich, W.~Zhang, L.~Sonderhouse, J.~M. Robinson, J.~Ye, F.~Riehle, and U.~Sterr, ``{1.5 $\mu$m Lasers with Sub-10 mHz Linewidth},'' \emph{Physical Review Letters}, vol. 118, 263202, 2017.}

\bibitem{Cherfan_2020oe}
\href{https://doi.org/10.1364/OE.28.000494}{C.~Cherfan, I.~Manai, S.~Zemmouri, J.-C. Garreau, J.-F. Cl{\'{e}}ment, P.~Szriftgiser, and R.~Chicireanu, ``{Acetylene-based frequency stabilization of a laser system for potassium laser cooling},'' \emph{Optics Express}, vol.~28, no.~1, pp. 494--502, 2020.}

\bibitem{Zhang_2020lpr}
\href{https://doi.org/10.1002/lpor.201900293}{W.~Zhang, L.~Stern, D.~Carlson, D.~Bopp, Z.~Newman, S.~Kang, J.~Kitching, and S.~B. Papp, ``Ultranarrow linewidth photonic‐atomic laser,'' \emph{Laser \& Photonics Reviews}, vol.~14, 1900293, 2020.}

\bibitem{numata_2010OE}
\href{https://doi.org/10.1364/OE.18.022781}{K.~Numata, J.~Camp, M.~A. Krainak, and L.~Stolpner, ``Performance of planar-waveguide external cavity laser for precision measurements,'' \emph{Opt. Express}, vol.~18, no.~22, pp. 22\,781--22\,788, 2010.}

\bibitem{Liu_2021optica}
\href{https://doi.org/10.1364/OPTICA.451635}{K.~Liu, N.~Chauhan, J.~Wang, A.~Isichenko, G.~M. Brodnik, P.~A. Morton, R.~O. Behunin, S.~B. Papp, and D.~J. Blumenthal, ``{36 Hz integral linewidth laser based on a photonic integrated 4.0 m coil resonator},'' \emph{Optica}, vol.~9, no.~7, pp. 770--775, 2022.}

\bibitem{Guo_2022sa}
\href{https://doi.org/10.1126/sciadv.abp9006}{J.~Guo, C.~A. McLemore, C.~Xiang, D.~Lee, L.~Wu, W.~Jin, M.~Kelleher, N.~Jin, D.~Mason, L.~Chang, A.~Feshali, M.~Paniccia, P.~T. Rakich, K.~J. Vahala, S.~A. Diddams, F.~Quinlan, and J.~E. Bowers, ``{Chip-based laser with 1-hertz integrated linewidth},'' \emph{Science Advances}, vol.~8, no.~43, eabp9006, 2022.}

\bibitem{Guo_2022apl}
\href{https://doi.org/10.1063/5.0088119}{Y.~Guo, X.~Li, M.~Jin, L.~Lu, J.~Xie, J.~Chen, and L.~Zhou, ``Hybrid integrated external cavity laser with a 172-nm tuning range,'' \emph{APL Photonics}, vol.~7, 066101, 2022.}

\bibitem{VanRees_2020oe}
\href{https://doi.org/10.1364/OE.386356}{A.~van Rees, Y.~Fan, D.~Geskus, E.~Klein, R.~Oldenbeuving, P.~van~der Slot, and K.-J. Boller, ``{Ring resonator enhanced mode-hop-free wavelength tuning of an integrated extended-cavity laser},'' \emph{Optics Express}, vol.~28, no.~4, pp. 5669--5683, 2020.}

\bibitem{Stern_2020ol}
\href{https://doi.org/10.1364/OL.398845}{L.~Stern, W.~Zhang, L.~Chang, J.~Guo, C.~Xiang, M.~A. Tran, D.~Huang, J.~D. Peters, D.~Kinghorn, J.~E. Bowers, and S.~B. Papp, ``{Ultra-precise optical-frequency stabilization with heterogeneous III–V/Si lasers},'' \emph{Optics Letters}, vol.~45, no.~18, pp. 5275--5278, 2020.}

\bibitem{fan_2016pj}
\href{https://doi.org/10.1109/JPHOT.2016.2633402}{Y.~Fan, J.~P. Epping, R.~M. Oldenbeuving, C.~G.~H. Roeloffzen, M.~Hoekman, R.~Dekker, R.~G. Heideman, P.~J.~M. van~der Slot, and K.-J. Boller, ``Optically integrated {InP-Si}$_3${N}$_4$ hybrid laser,'' \emph{IEEE Photonics Journal}, vol.~8, no.~6, 1505111, 2016.}

\bibitem{roeloffzen_2018jstqe}
\href{https://doi.org/10.1109/JSTQE.2018.2793945}{C.~G.~H. {Roeloffzen}, M.~{Hoekman}, E.~J. {Klein}, L.~S. {Wevers}, R.~B. {Timens}, D.~{Marchenko}, D.~{Geskus}, R.~{Dekker}, A.~{Alippi}, R.~{Grootjans}, A.~{van Rees}, R.~M. {Oldenbeuving}, J.~P. {Epping}, R.~G. {Heideman}, K.~W{\"o}rhoff, A.~{Leinse}, D.~{Geuzebroek}, E.~{Schreuder}, P.~W.~L. {van Dijk}, I.~{Visscher}, C.~{Taddei}, Y.~{Fan}, C.~{Taballione}, Y.~{Liu}, D.~{Marpaung}, L.~{Zhuang}, M.~{Benelajla}, and K.-J. {Boller}, ``Low-loss {Si}$_3${N}$_4$ {TriPleX} optical waveguides: Technology and applications overview,'' \emph{IEEE Journal of Selected Topics in Quantum Electronics}, vol.~24, no.~4, 4400321, 2018.}

\bibitem{liu_2001apl}
\href{https://doi.org/10.1063/1.1420585}{B.~Liu, A.~Shakouri, and J.~E. Bowers, ``Passive microring-resonator-coupled lasers,'' \emph{Applied Physics Letters}, vol.~79, no.~22, pp. 3561--3563, 2001.}

\bibitem{winkler2021frequency}
\href{https://photonics-benelux.org/wp-content/uploads/pb-files/proceedings/2021/Lasers/Lasers_04.pdf}{L.~V. Winkler, A.~van Rees, P.~J.~M. van~der Slot, C.~N{\"o}lleke, and K.-J. Boller, ``{Frequency stabilization of a hybrid-integrated InP-Si$_3$N$_4$ diode laser by locking to a fiber ring resonator},'' in \emph{25th Annual Symposium of the IEEE Photonics Benelux Chapter}, 2021.}

\bibitem{Kefelian_ol2009}
\href{https://doi.org/10.1364/OL.34.000914}{F.~Kéfélian, H.~Jiang, P.~Lemonde, and G.~Santarelli, ``Ultralow-frequency-noise stabilization of a laser by locking to an optical fiber-delay line,'' \emph{Optics Letters}, vol.~34, pp. 914--916, 2009.}

\bibitem{Day_1992jqe}
\href{https://doi.org/10.1109/3.135234}{T.~Day, E.~Gustafson, and R.~Byer, ``Sub-hertz relative frequency stabilization of two-diode laser-pumped Nd:YAQ lasers locked to a Fabry-Perot interferometer,'' \emph{IEEE Journal of Quantum Electronics}, vol.~28, pp. 1106--1117, 1992.}

\bibitem{Chauhan_2021nc}
\href{https://doi.org/10.1038/s41467-021-24926-8}{N.~Chauhan, A.~Isichenko, K.~Liu, J.~Wang, Q.~Zhao, R.~O. Behunin, P.~T. Rakich, A.~M. Jayich, C.~Fertig, C.~W. Hoyt, and D.~J. Blumenthal, ``{Visible light photonic integrated Brillouin laser},'' \emph{Nature Communications}, vol.~12, 4685, 2021.}

\bibitem{Yuan_2022oe}
\href{https://doi.org/10.1364/OE.458109}{Z.~Yuan, H.~Wang, P.~Liu, B.~Li, B.~Shen, M.~Gao, L.~Chang, W.~Jin, A.~Feshali, M.~Paniccia, J.~Bowers, and K.~Vahala, ``Correlated self-heterodyne method for ultra-low-noise laser linewidth measurements,'' \emph{Optics Express}, vol.~30, pp. 25\,147--25\,161, 2022.}

\bibitem{rutman1978characterization}
\href{https://doi.org/10.1109/PROC.1978.11080}{J.~Rutman, ``Characterization of phase and frequency instabilities in precision frequency sources: Fifteen years of progress,'' \emph{Proceedings of the IEEE}, vol.~66, no.~9, pp. 1048--1075, 1978.}

\bibitem{Swann_2000josab}
\href{https://doi.org/10.1364/JOSAB.17.001263}{W.~C. Swann and S.~L. Gilbert, ``{Pressure-induced shift and broadening of 1510–1540-nm acetylene wavelength calibration lines},'' \emph{Journal of the Optical Society of America B}, vol.~17, no.~7, pp. 1263--1270, 2000.}

\bibitem{Bjorklund_1983apb}
\href{https://doi.org/10.1007/BF00688820}{G.~C. Bjorklund, M.~D. Levenson, W.~Lenth, and C.~Ortiz, ``{Frequency modulation (FM) spectroscopy},'' \emph{Applied Physics B Photophysics and Laser Chemistry}, vol.~32, pp. 145--152, 1983.}

\bibitem{deLabachelerie_1995ol}
\href{https://doi.org/10.1364/OL.20.000572}{M.~de~Labachelerie, K.~Nakagawa, Y.~Awaji, and M.~Ohtsu, ``High-frequency-stability laser at 1.5 $\mu$m using {D}oppler-free molecular lines,'' \emph{Optics Letters}, vol.~20, no.~6, pp. 572--574, 1995.}

\bibitem{goldenstein_2017jqsrt}
\href{https://doi.org/10.1016/j.jqsrt.2017.06.007}{C.~S. Goldenstein, V.~A. Miller, R.~{Mitchell Spearrin}, and C.~L. Strand, ``{SpectraPlot.com: Integrated spectroscopic modeling of atomic and molecular gases},'' \emph{Journal of Quantitative Spectroscopy and Radiative Transfer}, vol. 200, pp. 249--257, 2017.}

\bibitem{Saleh_2015ao}
\href{https://doi.org/10.1364/AO.54.009446}{K.~Saleh, J.~Millo, A.~Didier, Y.~Kersal{\'{e}}, and C.~Lacro{\^{u}}te, ``{Frequency stability of a wavelength meter and applications to laser frequency stabilization},'' \emph{Applied Optics}, vol.~54, no.~32, pp. 9446--9449, 2015.}

\bibitem{konig2020performance}
\href{https://doi.org/10.1007/s00340-020-07433-4}{K.~K{\"o}nig, P.~Imgram, J.~Kr{\"a}mer, B.~Maa{\ss}, K.~Mohr, T.~Ratajczyk, F.~Sommer, and W.~N{\"o}rtersh{\"a}user, ``On the performance of wavelength meters: Part 2—frequency-comb based characterization for more accurate absolute wavelength determinations,'' \emph{Applied Physics B}, vol. 126, no.~5, 86, 2020.}

\bibitem{GonzalezGuerrero_2022jlt}
\href{https://doi.org/10.1109/JLT.2022.3171080}{L.~Gonzalez-Guerrero, R.~Guzman, M.~Ali, A.~Zarzuelo, J.~Cesar, D.~Dass, C.~Browning, L.~P. Barry, I.~Visscher, R.~Grootjans, C.~G.~H. Roeloffzen, and G.~Carpintero, ``{Injection Locking Properties of an InP-Si$_3$N$_4$ Dual Laser Source for Mm-wave Communications},'' \emph{Journal of Lightwave Technology}, vol.~40, no.~20, pp. 6685--6692, 2022.}

\bibitem{couturier2018laser}
\href{https://doi.org/10.1063/1.5025537}{L.~Couturier, I.~Nosske, F.~Hu, C.~Tan, C.~Qiao, Y.~Jiang, P.~Chen, and M.~Weidem{\"u}ller, ``Laser frequency stabilization using a commercial wavelength meter,'' \emph{Review of Scientific Instruments}, vol.~89, no.~4, 043103, 2018.}

\bibitem{Sakai_1991ptl}
\href{https://doi.org/10.1109/68.93244}{Y.~Sakai, I.~Yokohama, T.~Kominato, and S.~Sudo, ``Frequency stabilization of laser diode using a frequency-locked ring resonator to acetylene gas absorption lines,'' \emph{IEEE Photonics Technology Letters}, vol.~3, pp. 868--870, 1991.}

\bibitem{Lee_2013nc}
\href{https://doi.org/10.1038/ncomms3468}{H.~Lee, M.-G. Suh, T.~Chen, J.~Li, S.~A. Diddams, and K.~J. Vahala, ``Spiral resonators for on-chip laser frequency stabilization,'' \emph{Nature Communications}, vol.~4, 2468, 2013.}

\bibitem{Zhao_2021opt}
\href{https://doi.org/10.1364/OPTICA.432194}{Q.~Zhao, M.~W. Harrington, A.~Isichenko, K.~Liu, R.~O. Behunin, S.~B. Papp, P.~T. Rakich, C.~W. Hoyt, C.~Fertig, and D.~J. Blumenthal, ``{Integrated reference cavity with dual-mode optical thermometry for frequency correction},'' \emph{Optica}, vol.~8, no.~11, pp. 1481--1487, 2021.}

\bibitem{Hummon_2018opt}
\href{https://doi.org/10.1364/OPTICA.5.000443}{M.~T. Hummon, S.~Kang, D.~G. Bopp, Q.~Li, D.~A. Westly, S.~Kim, C.~Fredrick, S.~A. Diddams, K.~Srinivasan, V.~Aksyuk, and J.~E. Kitching, ``{Photonic chip for laser stabilization to an atomic vapor with $10^{-11}$ instability},'' \emph{Optica}, vol.~5, no.~4, pp. 443--449, 2018.}

\bibitem{Zektzer_2020lpr}
\href{https://doi.org/10.1002/lpor.201900414}{R.~Zektzer, M.~T. Hummon, L.~Stern, Y.~Sebbag, Y.~Barash, N.~Mazurski, J.~Kitching, and U.~Levy, ``{A Chip‐Scale Optical Frequency Reference for the Telecommunication Band Based on Acetylene},'' \emph{Laser {\&} Photonics Reviews}, vol.~14, no.~6, 1900414, 2020.}

\bibitem{Bovington_2014oe}
\href{https://doi.org/10.1364/OL.39.006017}{J.~T. Bovington, M.~J.~R. Heck, and J.~E. Bowers, ``Heterogeneous lasers and coupling to {Si$_3$N$_4$} near 1060 nm,'' \emph{Opt. Lett.}, vol.~39, no.~20, pp. 6017--6020, 2014.}

\bibitem{Tran_2022nat}
\href{https://doi.org/10.1038/s41586-022-05119-9}{M.~A. Tran, C.~Zhang, T.~J. Morin, L.~Chang, S.~Barik, Z.~Yuan, W.~Lee, G.~Kim, A.~Malik, Z.~Zhang, J.~Guo, H.~Wang, B.~Shen, L.~Wu, K.~Vahala, J.~E. Bowers, H.~Park, and T.~Komljenovic, ``Extending the spectrum of fully integrated photonics to submicrometre wavelengths,'' \emph{Nature}, vol. 610, pp. 54--60, 2022.}

\bibitem{Franken_2021ol}
\href{https://doi.org/10.1364/OL.433636}{C.~A.~A. Franken, A.~van Rees, L.~V. Winkler, Y.~Fan, D.~Geskus, R.~Dekker, D.~H. Geuzebroek, C.~Fallnich, P.~J.~M. van~der Slot, and K.-J. Boller, ``{Hybrid-integrated diode laser in the visible spectral range},'' \emph{Optics Letters}, vol.~46, no.~19, pp. 4904--4907, 2021.}

\bibitem{Winkler_2023spie}
\href{https://doi.org/10.1117/12.2667809}{L.~V. Winkler, K.~Gerritsma, A.~van Rees, P.~P.~J. Schrinner, M.~Hoekman, R.~Dekker, P.~J.~M. van~der Slot, C.~N{\"o}lleke, and K.-J. Boller, ``Silicon nitride hybrid-integrated diode laser at 637 nm,'' in \emph{Integrated Optics: Devices, Materials, and Technologies XXVII}, vol. 12424.\hskip 1em plus 0.5em minus 0.4em\relax SPIE Photonics West, 2023, {124241K}.}

\bibitem{Franken_2023arxiv}
\href{https://doi.org/10.48550/arXiv.2302.11492}{C.~A.~A. Franken, W.~A. P.~M. Hendriks, L.~V. Winkler, M.~Dijkstra, A.~R. do~Nascimento, A.~van Rees, M.~R.~S. Mardani, R.~Dekker, J.~van Kerkhof, P.~J.~M. van~der Slot, S.~M. García-Blanco, and K.~J. Boller, ``{Hybrid integrated near UV lasers using the deep-UV Al$_2$O$_3$ platform},'' \emph{arXiv}, 2302.11492, 2023.}

\end{thebibliography}

\end{document}